# Direct reconstruction of tissue conductivity with deconvolution in magneto-acousto-electrical tomography (MAET): theory and numerical simulation

Tong Sun[1], Dingqian Deng[1], Linguo Yu[1], Yi Chen[1], Chien Ting Chin[1], Mian Chen[1], Chunqi Chang[1], Siping Chen[1], Haoming Lin[1,a)], and Xin Chen[1,a)]

**AFFILIATIONS**

[1] Guangdong Provincial Key Laboratory of Biomedical Measurements and Ultrasound Imaging, National-Regional Key Technology Engineering Laboratory for Medical Ultrasound, School of Biomedical Engineering, Health Science Centre, Shenzhen University, Shenzhen 518060, China

a) Author to whom correspondence should be addressed: hm_lin@szu.edu.cn and chenxin@szu.edu.cn

**ABSTRACT**

Magneto-acousto-electrical tomography (MAET), a combination of ultrasound imaging and electrical impedance tomography (EIT), offers both high resolution (in comparison to EIT) and high contrast (in comparison to ultrasound imaging). It is used to map the internal conductivity distribution of an imaging object. However, conductivity reconstruction in MAET is a challenge, so conventional MAET is mainly devoted to mapping the conductivity interface. This is primarily because integration by parts is used in the theory derivation, and the simplified measurement formula suggests the voltage is proportional to the conductivity gradient, which leads to an error in the measurement formula. In this study, the measurement signal is expressed as the convolution of acoustic velocity and conductivity distribution without using integration by parts, which retains the low-frequency term in the measurement signal. Based on the convolution formula, we subsequently propose a direct conductivity reconstruction scheme with deconvolution by utilizing the low-frequency component. We verify the proposed method based on two two-dimension models and quantify the $L^2$ errors of reconstructed conductivity. Besides, we analyze factors influencing the reconstructed accuracy such as reconstructed regularization parameter, ultrasound frequency, and noise. We also demonstrate that the spatial resolution is not influenced by the duration of excitation ultrasound. With the contributions of the proposed method, conductivity imaging appears to be feasible for application to the early diagnosis in the future.

# I. INTRODUCTION

The electrical impedance of a biological tissue reflects the pathological and physiological characteristics of the tissue [1-3]. It has been reported that the dielectric properties of malignant breast tissue are much higher than those of normal breast tissue [4]. Electrical impedance tomography (EIT), which has great prospects in clinical applications due to its high imaging contrast, high-speed, low cost, and safety, aims to reconstruct the conductivity distribution of an object from electric potential measurements on the surface of the object [5-7]. However, low spatial resolution is a problem in EIT, especially for the interior regions of an object because the data measured on the surface are less sensitive to the variations of electrical properties in the interior. To improve the spatial resolution for interior regions, EIT has been combined with magnetic resonance imaging (MRI) [8, 9] or ultrasound [10-12].

In magnetic resonance electrical impedance tomography (MREIT), two pairs of external current are injected into the object through the surface electrodes [13-16]. Local magnetic field perturbations due to local conductivity perturbations are included in the MRI phase image. However, the high-resolution conductivity imaging in MREIT demands current injections of a few milliamperes, which may cause adverse effects of nerve and muscle stimulations depending on a chosen imaging area [17]. Besides, magnetic resonance electrical properties tomography (MREPT) is a promising technique to map the distribution of electrical properties (electrical conductivity and permittivity) inside an object at radio frequencies [18-22].

Regarding EIT combined with the ultrasound, magneto-acoustic tomography with magnetic induction (MAT-MI) [23-25] and magneto-acousto-electrical tomography (MAET) [26-28] are two representative methods. They are reciprocal methods proposed to reconstruct conductivity in the isotropic and anisotropic cases [29]. MAET [30], also called Lorentz force electrical impedance tomography [31] or Hall effect imaging [32], combines the ultrasound imaging and electrical impedance tomography, aiming to map the conductivity distribution inside the object. However, MAET has not been a robust method to image the conductivity distribution of object so far, because the magneto-acousto-electrical (MAE) signal was directly used to map the conductivity interface of imaging object rather than conductivity distribution in the conventional MAET [26-28, 32-36]. Although the electrical information needed to recover the conductivity distribution inside the object is included in the MAE signal, the recovery process remains a challenge in MAET.

We briefly review the development of MAET in previous studies and try to find the essential reason why it is difficult to reconstruct the conductivity distribution. Wen *et al*. [32] proposed MAET, called Hall effect imaging at the time, and built its theoretical basis. In their study, the measurement formula suggested that the measured signal was proportional to the gradient of conductivity. They inferred that the MAE signal appeared at the interface of conductivity. Later, the studies of MAET were roughly divided into two categories. Studies in the first category continued to adopt the theory and experimental method of Wen *et al*. [32]. For example, some researchers used the profile of the MAE signal to image the interface of conductivity with a higher spatial resolution [27, 28, 33]. Studies in the second category attempted to reconstruct the conductivity distribution from the conductivity gradient and subsequent the conductivity distribution by using a new algorithm [37]. They diverged further from the theory of Wen *et al*. [32]. For example, Zhou *et al* [37] proposed a scheme to reconstruct conductivity in one dimension with deconvolution in a simulation and phantom experiment. The measurement signal was deduced as the convolution of acoustic pressure and conductivity gradient by using integration by parts and ignoring the low-frequency term. Then, the conductivity gradient was recovered with deconvolution, and subsequent line integrals were used to obtain the conductivity distribution. However, the deconvoluted signal was not a standard gradient signal and line integrals could not be applied to the deconvoluted signal directly. To overcome this problem, they manually introduced Dirac's delta function by combining the polarity and amplitude of deconvoluted signal, and used line integrals to recover the conductivity distribution.

In conclusion, the theoretical basis of MAET originated from the theory established by Wen *et al*. [32]. However, the shortcoming of the theory lies in the use of acoustic pressure and integration by parts. The basic formula is $\mathbf{v} \times \mathbf{B}_0$, and the motion equation $\rho_0 (d\mathbf{v}/dt) = -\nabla p$ is used to substitute acoustic pressure for acoustic velocity. Later, integration by parts is used and the low-frequency term is ignored to obtain the measurement formula in the form of conductivity gradient. Theoretically, however, the substitution is unnecessary and not only increases the complexity but also leads to the misunderstanding of the measurement formula.

In this study, we derive the measurement formula in MAET with a concise form and explore the direct reconstruction of conductivity through the theoretical and numerical study. The following contributions are made to the ongoing MAET studies:

• In the theory part, the measurement signal is arranged as the convolution of acoustic velocity and conductivity distribution without using integration by parts, and thus the MAE signal including low-frequency component is obtained.

• A non-ideal conductivity model is proposed, and the results agree well with experimental results in different frequencies in other studies [31, 36, 37].

• The conductivity distribution can be directly reconstructed using deconvolution based on the theory and numerical simulation.

## II. Theory

Fig. 1(a) illustrates the schematic diagram of the MAE measurement system. We assume that the magnetic field is along the $x$ direction, and the sound wave only contains a component in the $z$ direction. Vector equation can be given as:

$$\boldsymbol{J}_L = \sigma \boldsymbol{v} \times \boldsymbol{B}_0 = \sigma v_z B_0 \hat{y}, \qquad (1)$$

where $\sigma$, $B_0$, $v_z$, and $\hat{y}$ are the conductivity, density of magnetic flux along the $x$ axis, vibrating velocity of ultrasound, and unit vector of the $y$ direction, respectively. We assume that the area of the ultrasound beam is $A$ and the width of the electrode along the $x$ direction is $W$. By introducing the electrical collection factor $\alpha$ of the experimental system, the current is expressed as:

$$I_y(t) = \frac{\alpha}{A} \int_W \int_y \int_z \sigma(y,z) v_z(y,z,t) B_0 dS dy. \qquad (2)$$

**A. The theory in conventional MAET**

In conventional MAET, the equation (2) was simplified

$$I_y(t) = \alpha \int_W \int_z \sigma(z) v_z(z,t) B_0 dS. \qquad (3)$$

The motion equation $\rho_0 (d\boldsymbol{v}/dt) = -\nabla p$ is expressed as:

$$v_z(z,t) = -\frac{1}{\rho_0} \int_{-\infty}^{t} \frac{\partial p(z,\tau)}{\partial z} d\tau. \qquad (4)$$

Substituting (4) into (3)

$$I_y(t) = -\frac{\alpha W B_0}{\rho_0} \int_{z_1}^{z_2} \sigma(z) \int_{-\infty}^{t} \frac{\partial p(z,\tau)}{\partial z} d\tau dz. \qquad (5)$$

According to integration by parts

$$I_y(t) = -\frac{\alpha W B_0}{\rho_0} \left[ \sigma(z) \int_{-\infty}^{t} p(z,\tau) d\tau \right]_{z_1}^{z_2} + \frac{\alpha W B_0}{\rho_0} \int_{z_1}^{z_2} \frac{\partial \sigma(z)}{\partial z} \int_{-\infty}^{t} p(z,\tau) d\tau dz. \qquad (6)$$

In some studies [36, 37], the first term on the right-hand side of (6) is regarded as the DC component or the term $\int_{-\infty}^{t} p(z,\tau) d\tau = 0$. Thus, the first term on the right-hand side is ignored, and formula (6) is simplified as

$$V(t) = \frac{\alpha W R_E B_0}{\rho_0} \int_{z_1}^{z_2} \frac{\partial \sigma(z)}{\partial z} \int_{-\infty}^{t} p(z,\tau) d\tau dz, \qquad (7)$$

where $R_E = \int_{z_1}^{z_2} (1/\sigma(z)) dz$ is the equivalent resistance of the entire model [37]. The sound wave is progressive in the measurement region, so the particle velocity is a function of *t-z/c*, where *c* is the propagation speed of the sound wave in the medium. According to (7)

$$V(t) = \frac{\alpha W R_E B_0}{\rho_0} \int_{z_1}^{z_2} \frac{\partial \sigma(z)}{\partial z} \int_{-\infty}^{t} p\left(\tau - \frac{z}{c}\right) d\tau dz \xrightarrow{z=c\tau} \varrho \int_{\tau_1}^{\tau_2} g(\tau) M(t-\tau) d\tau$$
$$= \varrho g(t) \otimes M(t), \qquad (8)$$

where $\varrho = \alpha W R_E B_0 c / \rho_0$, $g(\tau) := \partial \sigma(c\tau)/\partial c\tau$, and $M(t-\tau) := \int_{-\infty}^{t} p(\tau - z/c) d\tau$ represent a constant, the conductivity gradient function, and the ultrasound momentum, respectively. Please refer to other literatures [37, 38] for more details. According to (8), the induced voltage is equal to the convolution of conductivity gradient and the ultrasound momentum.

**B. The theory in proposed method**

In this section, we do not use integration by parts. Due to the progressive feature of the sound wave, we transform (2) into:

$$I_y(t) = \frac{\alpha W B_0}{A} \int_y \int_{z_1}^{z_2} \sigma(y,z) v_z(y, t-z/c) dz dy. \qquad (9)$$

Letting $z = c\tau$, we have

$$I_y(t) = \frac{\alpha c W B_0}{A} \int_y \int_{\tau_1}^{\tau_2} \sigma(y, c\tau) v_z(y, t-\tau) d\tau dy. \qquad (10)$$

Defining $h(\tau) := \sigma(c\tau)$, we write (10) as

$$V_y(t) = \eta \int_y \int_{\tau_1}^{\tau_2} h(y, \tau) v_z(y, t-\tau) d\tau dy = \eta \int_y h(y,t) \otimes v_z(y,t) dy, \qquad (11)$$

where $\eta = \alpha c W B_0 R_E / A$ is a constant for a specific experimental system, $R_E = \int_{y_1}^{y_2} \int_{z_1}^{z_2} (1/\sigma(y,z)) dy dz$ is the equivalent resistance of the entire model, and $\otimes$ is the convolution operator. According to (11), the MAE voltage is equal to the sum of convolution between conductivity and velocity in different *y* positions. Note that the acoustic velocity $v_z$ in (11) is a variable rather than the a constant (the sound speed of tissue is a constant). When the multi-layer object in Fig 2(a) is excited by the ultrasound, (11) can be simplified

$$V_y(t) = \eta h(t) \otimes \int_y v_z(y,t) dy. \qquad (12)$$

When the acoustic velocity is homogeneous along the *y* direction, (12) can be simplified the one-dimension (1D) model

$$V_y(t) = \eta' h(t) \otimes v_z(t), \qquad (13)$$

where $\eta' = \alpha c W B_0 R_E$ is a constant for a specific experimental system. Unlike almost all previous studies [26, 27, 31-33, 35-38], we do not use integration by parts in the theory derivation because we need to use the low-frequency component of the measurement signal. We assume that the acoustic velocity is uniform along *y* direction, the (11) is written as

$$V_y(t) = \eta \int_y h(y,t) dy \otimes v_z(t) = \eta' h^{av}(t) \otimes v_z(t), \qquad (14)$$

where $h^{av}(t) := 1/A \int_y h(y,t) dy$ is the average conductivity. In the case of the frequency response $R(t)$ of the transducer, the MAE voltage can be expressed as

$$V(t) = \eta' h^{av}(t) \otimes T(t) \otimes R(t), \tag{15}$$

where $T(t)$ is the transmitted signal (sinusoidal burst in this study), and $v_z(t) = T(t) \otimes R(t)$. For systems used in experiments, such as power amplifier and low noise amplifier, it is reasonable to consider their bandwidths, so we have

$$V(t) = \eta' h^{av}(t) \otimes T(t) \otimes R(t) \otimes E(t), \tag{16}$$

where $E(t)$ is the frequency response of the experimental system. Because the observation signal is usually corrupted by zero-mean additive Gaussian noise (AWGN) $\gamma(t)$, (16) can be expressed as:

$$V(t) = \eta' h^{av}(t) \otimes v_z(t) \otimes E(t) + \gamma(t). \tag{17}$$

For convenience, we ignore the $E(t)$ in the remainder of this paper. The discrete-time format of (17) is expressed as:

$$V(n) = \eta' h^{av}(n) \otimes v_z(n) + \gamma(n). \tag{18}$$

The deconvolution problem in MAET is described by estimating $h^{av}(n)$ from observation signal $V(n)$, known system $v_z(n)$ and unknown noise $\gamma(n)$.

We obtain a native deconvolution estimate $\tilde{h}^{av}(n)$ using the operator inverse $\mathcal{V}^{-1}$ as:

$$\tilde{h}^{av}(n) := \mathcal{V}^{-1} V(n)/\eta' = h^{av}(n) + \mathcal{V}^{-1} \gamma(n)/\eta'. \tag{19}$$

For the orthonormal basis $\{b_k\}_{k=0}^{N-1}$ for $\mathbb{R}^{N-1}$, the estimate $\tilde{h}^{av}(n)$ can be expressed as [39]:

$$\tilde{h}^{av} := \sum_{k=0}^{N-1} (\langle h^{av}, b_k \rangle + \langle \mathcal{V}^{-1} \gamma/\eta', b_k \rangle) b_k. \tag{20}$$

When the basis $\{b_k\}_{k=0}^{N-1}$ is Fourier basis, the Fourier coefficients of the noise $\gamma$ are significantly amplified. Consequently, a Fourier shrinkage $\lambda_k, 0 \leq \lambda_k \leq 1$ is introduced:

$$\tilde{h}_\lambda^{av} := \sum_{k=0}^{N-1} (\langle h^{av}, b_k \rangle + \langle \mathcal{V}^{-1} \gamma/\eta', b_k \rangle) \lambda_k b_k = h_\lambda^{av} + \mathcal{V}^{-1} \gamma_\lambda/\eta', \tag{21}$$

where $h_\lambda^{av}$ and $\mathcal{V}^{-1} \gamma_k$ are respectively the retained part and leaked part of $h^{av}$, and the shrinkage $\lambda_k$ can be interpreted as a form of the regularization in the deconvolution.

The Fourier domain is the traditional choice for deconvolution because the convolution simplifies to scalar Fourier operations. That is, (18) can be rewritten as:

$$V(f_k) = \eta' H^{av}(f_k) V_z(f_k) + \Gamma(f_k), \tag{22}$$

where $V(f_k)$, $H^{av}(f_k)$, $V_z(f_k)$, and $\Gamma(f_k)$ are the discrete Fourier transforms (DFTs) of $V(n)$, $h^{av}(n)$, $v_z(n)$, and $\gamma(n)$, respectively, and $f_k := \pi k/N, k = -N/2+1,...,N/2$ are the normalized DFT frequencies. The pseudo-inversion operation of (19) in the Fourier domain is written as:

$$\tilde{H}^{av}(f_k) := \begin{cases} H^{av}(f_k) + \dfrac{\Gamma(f_k)}{\eta' V_z(f_k)}, & \text{if } |V_z(f_k)| > 0 \\ 0 & \text{otherwise} \end{cases}, \tag{23}$$

where $\tilde{H}^{av}(f_k)$ is the DFT of $\tilde{h}^{av}(n)$. When $|V_z(f_k)| \approx 0$, the noise components are particularly amplified.

Deconvolution with Fourier shrinkage can attenuate the amplified noise in $\tilde{H}_{av}(f_k)$ with shrinkage

$$\lambda_k(f_k) = \frac{|V_z(f_k)|^2}{|V_z(f_k)|^2 + \Lambda(f_k)}, \tag{24}$$

where $\Lambda(f_k)$ represents regularization terms which control the amount of shrinkage. The DFT components of the estimate $\tilde{h}_\lambda^{av}$ are expressed as:

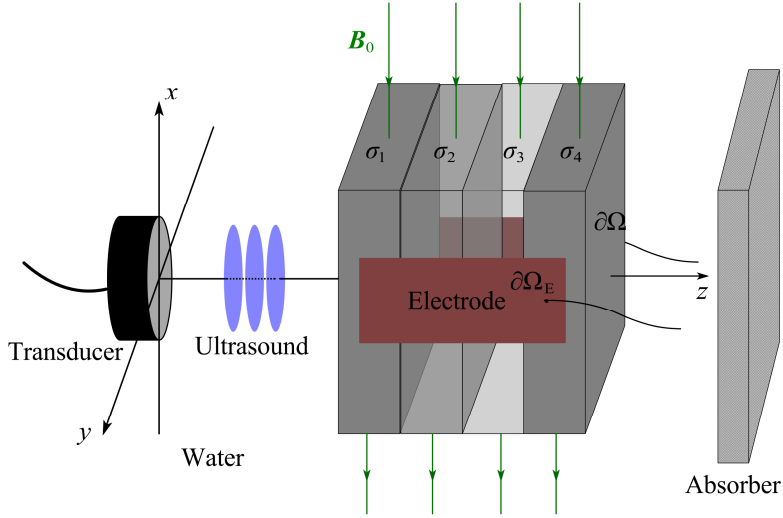

**FIG 1**. Schematic diagram of the MAE measurement system.

$$\begin{aligned}
\tilde{H}_\lambda^{av}(f_k) &:= \tilde{H}^{av}(f_k)\lambda_k(f_k) \\
&= H^{av}(f_k)\left(\frac{|V_z(f_k)|^2}{|V_z(f_k)|^2+\Lambda(f_k)}\right) + \frac{\Gamma(f_k)}{\eta' V_z(f_k)}\left(\frac{|V_z(f_k)|^2}{|V_z(f_k)|^2+\Lambda(f_k)}\right), \\
&= H_\lambda^{av}(f_k) + \frac{\Gamma_\lambda(f_k)}{\eta' V_z(f_k)}
\end{aligned} \qquad (25)$$

where $H_\lambda^{av}$ and $\Gamma_\lambda/V_z$ denote the DFTs of $h_\lambda^{av}$ and $\mathcal{V}^{-1}\gamma_k$, respectively. For different deconvolution techniques, such as linear time-invariant (LTI) Wiener deconvolution and Tikhonov-regularized deconvolution [40], the choice of shrinkage $\lambda$ is different. LTI Wiener deconvolution sets as defined as follows:

$$\lambda_k(f_k) = \frac{|V_z(f_k)|^2}{|V_z(f_k)|^2 + \alpha \frac{N\kappa^2}{|H^{av}(f_k)|^2}}, \qquad (26)$$

where $\alpha$ is a regularization parameter, and $\kappa$ denotes the variance of noise. The shrinkage $\lambda_k = 1$ hardly shrinks with $\alpha = 0$. In contrast, shrinkage $\lambda_k \approx 0$ shrinks more with $\alpha = 1$. Tikhonov-regularized deconvolution assumes a flat signal spectrum:

$$\lambda_k(f_k) = \frac{|V_z(f_k)|^2}{|V_z(f_k)|^2 + \chi}, \qquad (27)$$

with $\chi > 0$. The shrinkage $\lambda_k$ in (27) has actually already been used in MAET with the term of noise-to signal ratio (NSR) [38]. In this study, we also utilize (27) due to the lack of a priori information of signal and noise. Finally, we directly recover the estimated conductivity distribution $\tilde{h}_\lambda^{av}(n) = \tilde{\sigma}_\lambda^{av}(cn)$ via the inverse DFT of $\tilde{H}_\lambda^{av}(f_k)$.

## III. Methods

### A. Governing equation in MAET

In this section, we demonstrate the governing equations and boundary conditions used in COMSOL Multiphysics. A general geometry of MAET is shown in Fig. 1(a). With the static magnetic field $\boldsymbol{B}_0$ along the $x$ direction, the object with conductivity distribution $\sigma(y,z)$ is excited by the particle velocity $\boldsymbol{v}(y,z,t)$ along the $+z$ direction transmitted by one single-element transducer. Without considering the acoustic attenuation and viscosity, the governing equation is as follows:

$$\nabla \cdot (\sigma \nabla u) = \nabla \cdot (\sigma \boldsymbol{v} \times \boldsymbol{B}_0), \qquad (28)$$

where $u$ denotes the electric potential.

For the time explicit method, the governing wave equations are formulated as the first order system, terms of the linearized continuity equation and the linearized momentum equation, as

$$\frac{1}{\rho c^2}\frac{\partial p(z,t)}{\partial t} + \nabla \cdot \boldsymbol{v}(z,t) = 0$$
$$\rho \frac{\partial \boldsymbol{v}(z,t)}{\partial t} + \nabla p(z,t) = 0, \quad (29)$$

where $p, \rho$ and $c$ represent the acoustic pressure, the mass density of medium and the speed of sound, respectively. After the acoustic velocity $\boldsymbol{v}(z,t)$ is obtained, we substitute it into (28) to solve the electric potential $u$.

The boundary condition corresponding to areas without an electrode is given by:

$$\boldsymbol{J} \cdot \boldsymbol{n} = 0, \quad (30)$$

where $\boldsymbol{J}$ and $\boldsymbol{n}$ respectively denote the current density and out normal vector on the boundary. In contrast, for the boundary condition corresponding to areas with electrode, it is better to use a high-impedance receiver [41]:

$$\int_{\partial \Omega_E} -\boldsymbol{n} \cdot \boldsymbol{J} dS = 0. \quad (31)$$

For the acoustics module, we use the surface velocity boundary condition to produce the acoustic wave:

$$\boldsymbol{v}_t \cdot \boldsymbol{n} = -v_n(t), \quad (32)$$

where $\boldsymbol{v}_t$, $\boldsymbol{n}$, and $-v_n(t)$ are the vector velocity, outward unit normal vector, and velocity transmitted by the transducer, respectively.

**B. Simulation setup**

To obtain the simulated data, we use finite element software COMSOL Multiphysics (version 5.5, COMSOL Inc., Sweden) to build the geometry, material properties, physical field, mesh, and solver. For the geometry part, we build one continuous object with regions of different material properties rather than building separate objects (see supplemental material for details). Generally, if the material properties of tissues were different, one needed to build different geometry regions to define the material property in COMSOL. However, it is difficult to define exact transition region between several different geometry regions. Consequently, we define a position-dependent conductivity function (see section III.C). In the material part, $\sigma(z)$ (the position-dependent conductivity function) is filled in the 'conductivity' field in COMSOL. We use pressure acoustics, time explicit interface based on the Runge-Kutta discontinuous Galerkin (DG) method [42] to simulate the propagation of ultrasound considering some frequencies of ultrasound waves (above 2.00 MHz) are high [35], and we use the magnetic field module to model the magneto-acousto-electrical phenomenon.

In this study, we configure a multi-layer conductivity model denoted as Model I shown in Fig. 2(a) and a general two-dimension model denoted as Model II shown in Fig. 2(b). For Model I, we demonstrate the 1D characterization. Finally, the B-scan mode along $y$ direction (range from -24 mm to 24 mm with a step 2 mm) is adopted to obtain the MAE signals and the conductivity images are mapped for Model I and II.

**C. Non-ideal conductivity model versus ideal conductivity model**

In previous studies, the conductivity model in $z$ direction was nearly ideal (the transition band was extremely small). However, in practice, the conductivity interface changes continuously with a large transition band. Consequently, in this study, we propose a non-ideal conductivity model with a large transition band. The ideal conductivity model in $z$ direction was expressed as the weighted sum of jump function:

$$\sigma(z) = \sum_{i=1}^{M} W_i \varepsilon_I (z - z_i), \quad (33)$$

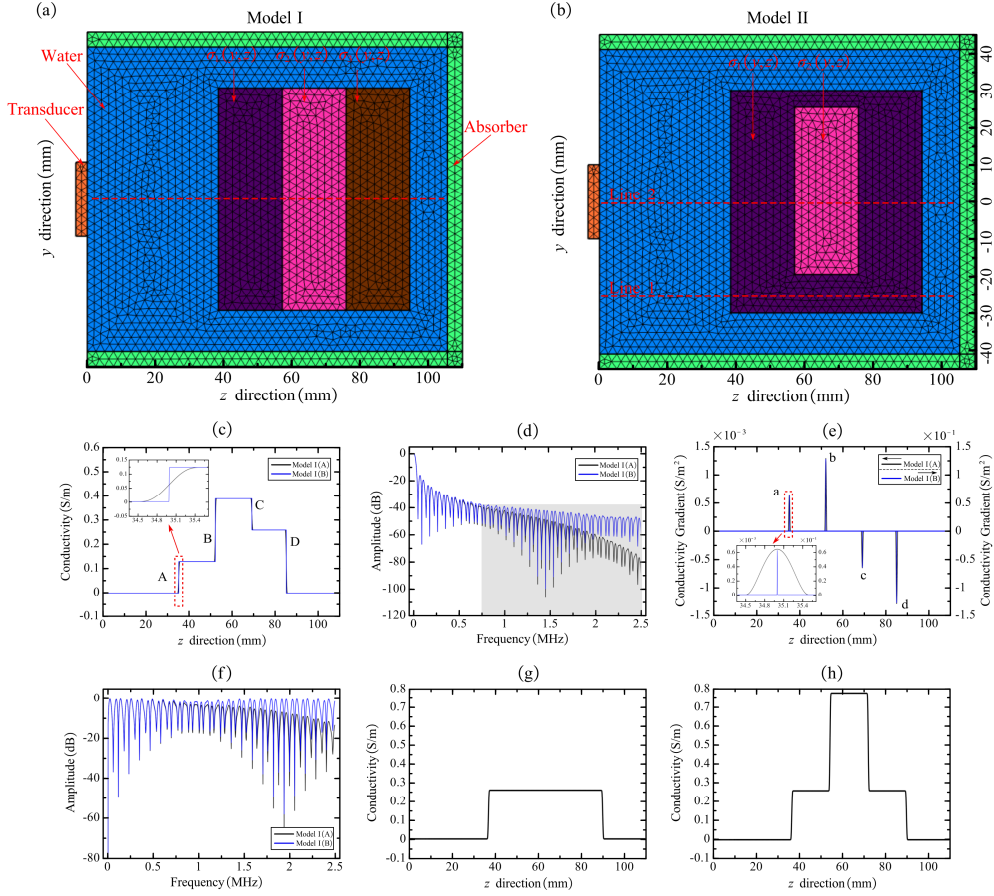

**FIG 2**. (a) The configuration of Model I, red dashed line denotes the conductivity profile of (c). (b) The configuration of Model II, red dashed lines denote the conductivity profile of (g) and (h), respectively. (c) The conductivity distribution of Model I(A) and Model I(B). (d) The frequency spectra of Model I(A) and Model I(B). (e) The gradient of Model I(A) and Model I(B). (f) The frequency spectra of conductivity gradient for Model I(A) and Model I(B). (g) The conductivity distribution corresponding to Line 1 of Model II. (h) The conductivity distribution corresponding to Line 2 of Model II

where $\varepsilon_I(z) = \begin{cases} 0, & z < 0 \\ 1, & z \geq 0 \end{cases}$ is the ideal jump function, $W_i$ is the weighted coefficient, $M$ is the number of conductivity variation, and $z_i$ is the position of jump. The only restriction we impose on the choice of weights $W_i$ is that their sum should be equal to zero, that is,

$$\sum_{i=1}^{M} W_i = 0. \tag{34}$$

For the non-ideal conductivity model, the conductivity variation is continuous

$$\varepsilon_N(z) = \begin{cases} 0, & z < 0 \\ kz, & 0 \leq z \leq 1/k, \\ 1, & z > 1 \end{cases} \tag{35}$$

where $k$ is the slope of the transition band. When the inverse of the slope $1/k$ is far larger than the wavelength of ultrasound, the conductivity model can be approximately regarded as the ideal conductivity model. In this study, we set $k$ 1e3 and 1e5 (corresponding to transition bands of 1 and 0.01 mm, respectively). When $k$ is equal to 1e5, the conductivity model is nearly ideal because the wavelength of ultrasound (about 0.5-10 mm) is far larger than the transition band (0.01 mm). The non-ideal and ideal conductivity models are denoted as Model I(A) and Model I(B), respectively.

We define the position-dependent conductivity function $\sigma(z)$ along the $z$ direction, shown in Fig. 2(c), using $\varepsilon_N(z)$ as follows:

$$\sigma(z) = [\varepsilon_N(z-35) + 2\varepsilon_N(z-52) - \varepsilon_N(z-69) - 2\varepsilon_N(z-86)] \times 0.1285, \ 0 \leq z \leq 110, \quad (36)$$

where 35, 52, 69, and 86 (unit: mm) correspond to the variation of conductivity with the position. The amplified information of the transition zone in the first conductivity variation is also shown in Fig. 2(c).

The frequency spectra of Model I(A) (black line) and Model I(B) (blue line) are presented in Fig. 2(d), showing that the differences between Model I(A) and Model I(B) are mainly above 0.75 MHz (gray region). The frequency spectra of Model I(B) are relatively flat, whereas that of Model I(A) decrease rapidly with increasing frequency. The difference in amplitude is about 18 dB for 2.20 MHz.

To observe the difference of time-domain waveform between Model I(A) and Model I(B), we use the excited signals of six-cycle sinusoidal burst in two frequencies (0.20 and 2.20 MHz) (see section IV.A).

**D. Effective frequency analysis**

We calculate the effective frequency $\bar{f}$ of Model I(A) in Fig. 2(c) by using the following formula:

$$\bar{f} = \frac{\sum_{i=1}^{L} f_i |H(f_i)|}{\sum_{i=1}^{L} |H(f_i)|}, \quad (37)$$

where $L$, $f_i$, and $|H(f_i)|$ respectively represent the length of Fourier transformation, the frequency component, and the frequency-domain absolute amplitude of the non-ideal conductivity model. We calculate the standard deviation of the effective frequency $f_{sd}$ by using the following formula:

$$f_{sd} = \sqrt{\frac{\sum_{i=1}^{L} (f_i - \bar{f})^2 |H(f_i)|}{\sum_{i=1}^{L} |H(f_i)|}}. \quad (38)$$

The frequency range is from 0 to 3.00 MHz. The resulting effective frequency is 0.209 ± 0.365 MHz. After obtaining the effective frequency and deviation, we choose the appropriate frequencies of ultrasound for the numerical study.

**E. Excited signal**

In this section, we set the excited frequencies of 0.10, 0.15, 0.20, 0.25, and 0.30 MHz around the effective frequency calculated in section III.D. To compare the difference of MAE voltage between low- and high-frequency, we also add a high-frequency 2.20 MHz. The 6-cycle sinusoidal burst at 0.20 and 2.20 MHz are presented in Fig. 3(a) and (c). The corresponding frequency spectra are shown in Fig 3(b) and (d). In addition, we also present the frequency spectra of conductivity distribution (blue line in Fig 3(b) and (d)).

To compare MAE voltages generated by conductivity distribution model and conductivity gradient model, we also use numerical software MATLAB to simulate the MAE voltage with the excited signal of 1.10 MHz 6-cycle sinusoidal burst. The reason why we use this frequency is that the induced signal in different conductivity position could be clearly observed. Other frequencies like 1.00 or 1.20 MHz are also applicable.

The COSMOL software is used to simulate all MAE signals in 0.10, 0.15, 0.20, 0.25, 0.30 and 2.20 MHz. For Model I, if there is no special identification, simulation is conducted based on Model I(A). All numerical computations are implemented with a Windows-based workstation with two Intel Xeon Gold 5118 2.30 GHz dual processors (12 cores) and 128 GB memory.

**F. Reconstructed method and evaluation**

After we obtain the MAE signal and acoustic velocity (both normalized to one), we apply DFT to these two signals. Then, we compute the shrinkage $\lambda_k(f_k)$ according to (27) using the acoustic velocity. Next, we obtain the estimate of conductivity distribution $\tilde{H}_\lambda(f_k)$ in the frequency-domain using (25). Then we directly recover the estimated conductivity distribution $\tilde{h}_\lambda(n) = \tilde{\sigma}_\lambda(cn)$ via the inverse DFT of $\tilde{H}_\lambda(f_k)$. Finally, we normalize the amplitude of the conductivity distribution to one.

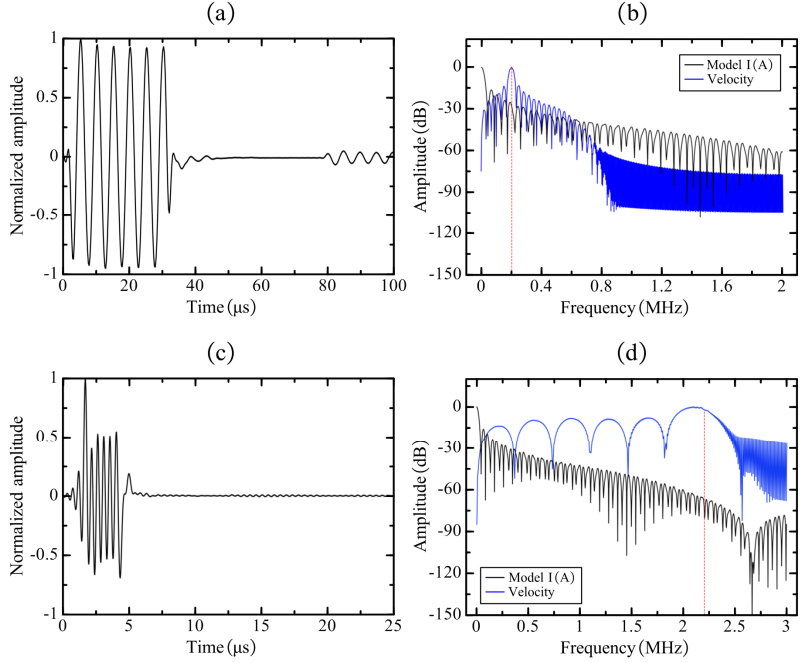

**FIG 3.** Time domain normalized acoustic velocity at: (a) 0.20 MHz and (c) 2.20 MHz. Frequency spectra of acoustic velocity (blue line) at (b) 0.20 MHz and (d) 2.20 MHz.

To quantify the reconstructed errors, we define the relative $L^2$-error:

$$E(\%) := \frac{\|\tilde{\sigma}_R - \sigma_O\|_2}{\|\sigma_O\|_2}, \tag{39}$$

where $\tilde{\sigma}_R$ and $\sigma_O$ represent the reconstructed and true conductivity, respectively.

To explore the effect of noise on the reconstruction, we add Gaussian white noise with mean zero and variance one to the simulation output of the MAE signal. We set the noise amplitude to 2 and 4 μV, separately.

## IV. Results

### A. Time- and frequency-domain MAE signals

Based on Model I(A) and Model I(B), the MAE voltages excited by ultrasound at frequencies of 0.20 and 2.20 MHz are presented in Fig. 4(a) and (c), respectively. As for the MAE voltage for 0.20 MHz, the wave packets induced by different conductivity variation experience aliasing. In addition, no significant difference is observed in the time-domain MAE voltage in Fig. 4(a). As for the time-domain MAE voltage for 2.20 MHz, four clear wave packets are induced, but the waveform is a little distorted. For convenience, we divide the MAE voltage into two components: the conductivity profile component (CPC) and the ultrasound induced component (UIC). CPC reflects the information of conductivity distribution, whereas UIC is solely related to the excited ultrasound. According to the detailed plots of first and third wave packets in Fig .4(c), the UIC based on Model I(A) is smaller than that of Model I(B), whereas the CPCs of the two models are almost the same.

The corresponding frequency spectra of MAE voltages are displayed in Fig. 4(b) and (d). For 0.20 MHz, there is no significant difference in the frequency spectrum. The wavelength of ultrasound is about 7.25 mm for 0.20 MHz, and the conductivity models with the transition bands of 0.01 and 1 mm are nearly equivalent for this wavelength. The frequency spectra of the two conductivity models are nearly equal for frequencies below 0.30 MHz, whereas the amplitude of the frequency spectrum for 2.20 MHz is different. The amplitude of the Model I(B) is larger (beyond 12 dB) than that of the Model I(A). In practice, the difference of the frequency spectrum is in accordance with that of the conductivity model. Note that the UIC dominates for 0.20 MHz, as shown in Fig. 4(b), whereas the CPC dominates for 2.20 MHz, as shown in Fig. 4(d), because the amplitude of the frequency spectrum for the conductivity distribution model decreases rapidly with increasing frequency.

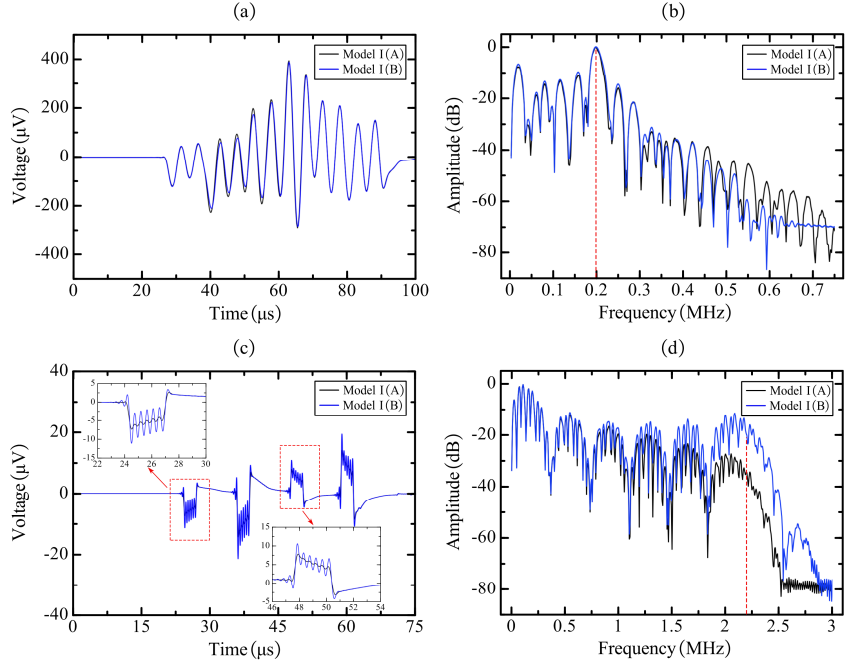

**FIG 4.** MAE voltages of Model I(A) and Model I(B) at (a) 0.20 MHz and (c) 2.20 MHz, and the frequency spectra of MAE voltages at (b) 0.20 MHz and (d) 2.20 MHz.

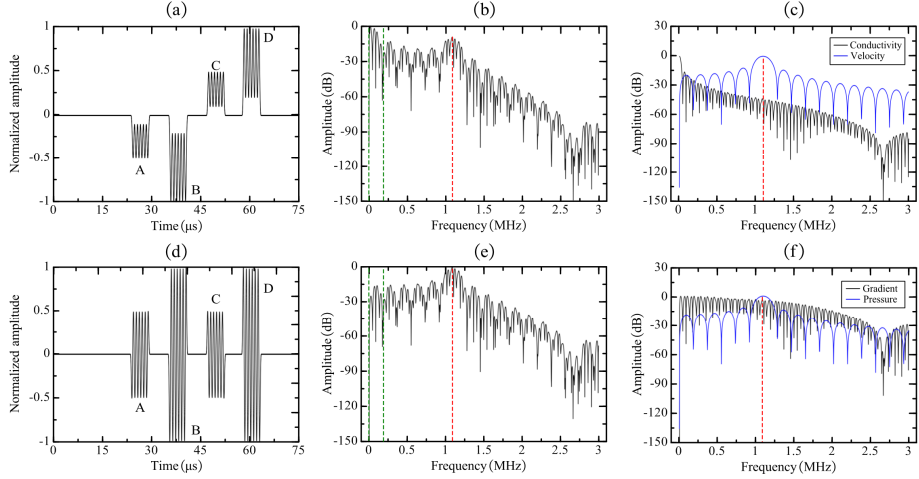

**FIG 5.** (a) The MAE voltage obtained by the convolution of conductivity distribution and acoustic velocity, (b) the frequency spectrum of (a), (c) the frequency spectra of conductivity distribution (black solid line) and acoustic velocity (blue solid line), (d) The MAE voltage obtained by the convolution of conductivity gradient and acoustic pressure, (e) the frequency spectrum of (d), the frequency spectrum of conductivity gradient (black solid line) and acoustic pressure (blue solid line).

### B. Conductivity distribution model versus conductivity gradient model

The MAE signals based on conductivity distribution and conductivity gradient model are presented in Fig. 5(a) and (d), respectively. Four wave packets (A, B, C, and D) are both clearly displayed in Fig. 5(a) and (d) at the conductivity interface. But the waveforms between them are a little different. In Fig .5(a), the wave packets are composed of CPC and UIC. The relative polarities of CPC in Fig. 5(a) are consistent with that of the conductivity, and in contrast CPC could not be observed in Fig. 5(d).

The corresponding frequency spectrum of MAE signals were displayed in Fig. 5(b) and (e), respectively. The main difference between them lied in the low-frequency ranging from 0 to 0.20 MHz (within the cyan dashed line). In Fig. 5(b), the CPC was comparable with UIC, and in contrast the CPC in Fig. 5(e) was far smaller than UIC. The reason can be explained by the frequency spectrums of conductivity distribution and conductivity gradient in Fig. 5(c) and (f).

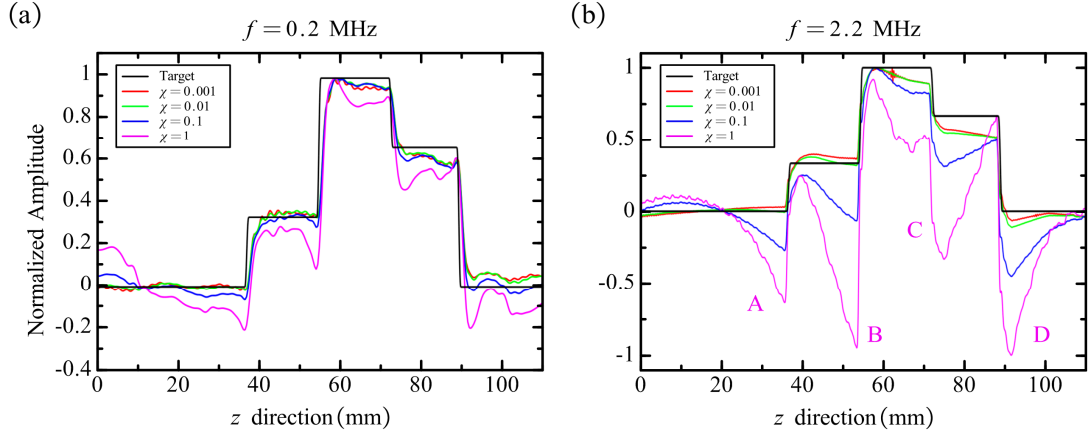

**FIG 6**. Reconstructed conductivity distribution by using different regularization parameters $\chi$ = 0.001 (red solid line), 0.01 (green line), 0.1 (blue line), and 1 (magenta line) for (a) 0.20 MHz and (b) 2.20 MHz.

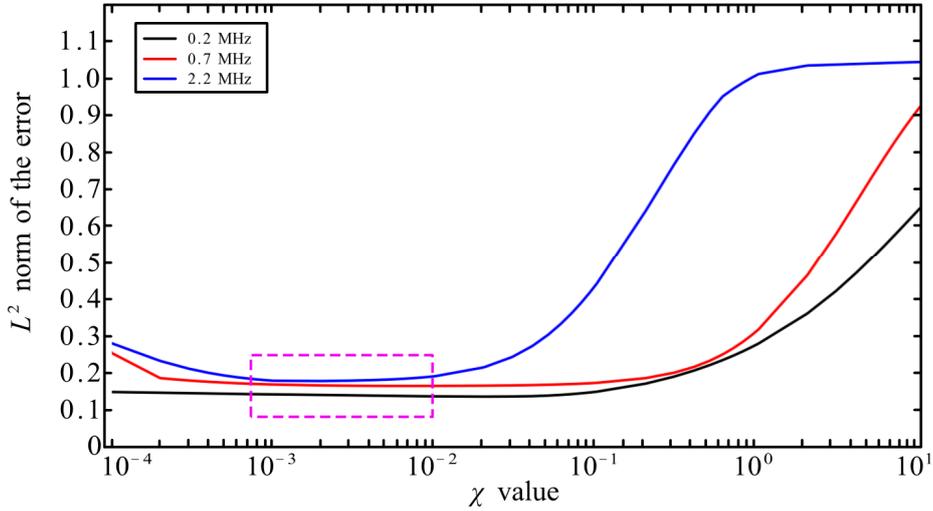

**FIG 7**. The $L^2$ norm of the error versus regularization parameter $\chi$ for excitation frequencies of 0.20 MHz, 0.70 MHz, and 2.20 MHz. The magenta dashed box represents the flat error with different $\chi$.

**C. Directly reconstructed conductivity distribution and error**

*1. The reconstructed conductivity and error versus regularization parameter $\chi$*

In this sub-section, we explore the reconstruction error variation by changing the regularization parameter $\chi$ from $10^{-4}$ to 10. The regularization parameter $\chi$ in (27) controls the shrinkage $\lambda_k$. When the value of $\lambda_k$ increases, the signal and noise are both suppressed. The reconstructed conductivity distributions with excitation frequencies of 0.20 and 2.20 MHz are shown in Fig. 6. In Fig. 6(a), the profiles of reconstructed conductivity in 0.20 MHz rarely vary when $\chi$ ranges from 0.001 to 0.1. When $\chi$ is equal to 1, the low- and high-frequency components of reconstructed conductivity are both compressed by the shrinkage, and the error also increases. In contrast, the shrinkage effect is obvious at 2.20 MHz. In Fig. 6(b), when $\chi$ is equal to 1, four clear peaks A, B, C, and D appear. Peaks A and B have the same polarity as do C and D, whereas the polarities of A and C are opposite.

We quantify the reconstructed error for the excitation frequencies of 0.20, 0.70, and 2.20 MHz with noise-free data, as shown in Fig. 7. In the entire range of the regularization parameter, the error of 0.20 MHz is the minimum, and the error of 2.20 MHz is the maximum. When $\chi$ is smaller than 0.001, the reconstructed conductivity is influenced by the small velocity component, which leads to large error. When $\chi$ was larger than 0.1, the signal and noise were both compressed by the shrinkage which led to large error (see Fig. 6(b)). Consequently, to obtain good results, an appropriate regularization parameter should be chosen. The magenta dashed box in Fig. 7 represents the flat error with different regularization parameters, so we set $\chi$ to 0.01 in the following reconstruction.

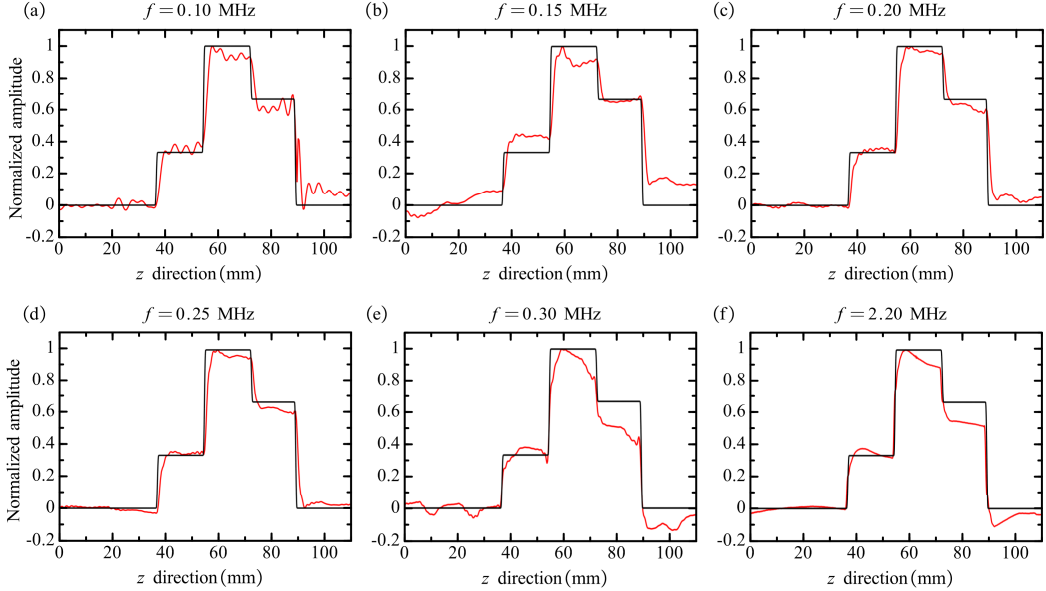

**FIG 8**. Reconstructed conductivity distribution of different ultrasound frequencies with noise-free data in (a) 0.10 MHz, (b) 0.15 MHz, (c) 0.20MHz, (d) 0.25 MHz, (e) 0.30 MHz, and (f) 2.20 MHz. Black and red lines respectively denote target and reconstructed conductivity.

*2. The effect of ultrasound frequency.*

The reconstructed conductivity distributions with different excitation frequencies are shown in Fig. 8. Black and red lines respectively denote the target and reconstructed conductivity. In Fig. 8(a), the reconstructed transition band is larger than the target, suggesting that a large amount of the high-frequency component is lost when the object is excited by the ultrasound at 0.10 MHz. In contrast, the transition bands in Fig. 8(e) and (f) are more consistent with the target. Thus, the reconstructed transition bands (composed by some high-frequency components) are related to the excited frequency. According to Fig. 8, the reconstructed conductivity distributions in different frequencies are similar with noise-free data. The reconstructed errors with noise-free data are listed in table 1 (with 0 $\mu$V-level noise). The maximum and minimax errors are respectively 0.2339 and 0.1289 for 0.30 and 0.25 MHz. One could find that the error for the frequency of 0.30 MHz is significantly larger than the errors for frequencies below 0.25 MHz.

To study the tendency of reconstructed error towards the excitation frequency, we simulate the MAE voltage from 0.05 to 2.80 MHz. To characterize inherent features, we use the optimal regularization parameter in each frequency. The error curve (red line) with noise-free data is presented in Fig. 9. The error from 0.05 to 2.80 MHz is stable, which means the reconstructed conductivity distribution is not influenced by the frequency in the presence of 0 $\mu$V noise.

*3. The effect of noise.*

When we use the data with 2 $\mu$V and 4 $\mu$V noise, the errors for frequencies below 0.30 MHz are similar, although the error increases gradually with increasing frequency. However, the error for 0.30 MHz is larger than that for other frequencies below 0.90 MHz and it can be regarded as an outlier. Moreover, when the frequency exceeds 2.00 MHz, the error first declines and later increases. Generally, errors for the frequencies around the effective frequencies are smaller than errors for other frequencies, which suggests the parameter effective frequency is reasonable.

As mentioned above, we set the regularization parameter $\chi$ to 0.01. The reconstructed conductivity distributions with different levels of noise are presented in Fig. 10. After adding noise, the reconstructed distributions for 0.10, 0.15, 0.20, and 0.25 MHz are basically consistent with the target. The reconstructed conductivity for 0.25 MHz is visually the best among these results. The reconstructed error for 0.25 MHz comes from the transition regions. In contrast the reconstructed distribution for 2.20 MHz is contaminated by noise, and the profile of reconstructed conductivity can hardly be recovered.

**Table 1.** $L^2$-errors of the reconstructed conductivity distribution.

| Noise (μV) /Frequency (MHz) | 0.10 | 0.15 | 0.20 | 0.25 | 0.30 | 2.20 |
|---|---|---|---|---|---|---|
| 0 μV | 0.1769 | 0.1985 | 0.1358 | 0.1289 | 0.2339 | 0.1898 |
| 2 μV | 0.1979 | 0.2023 | 0.1489 | 0.1418 | 0.2561 | 0.4421 |
| 4 μV | 0.2568 | 0.2114 | 0.1689 | 0.1632 | 0.2962 | 0.6304 |

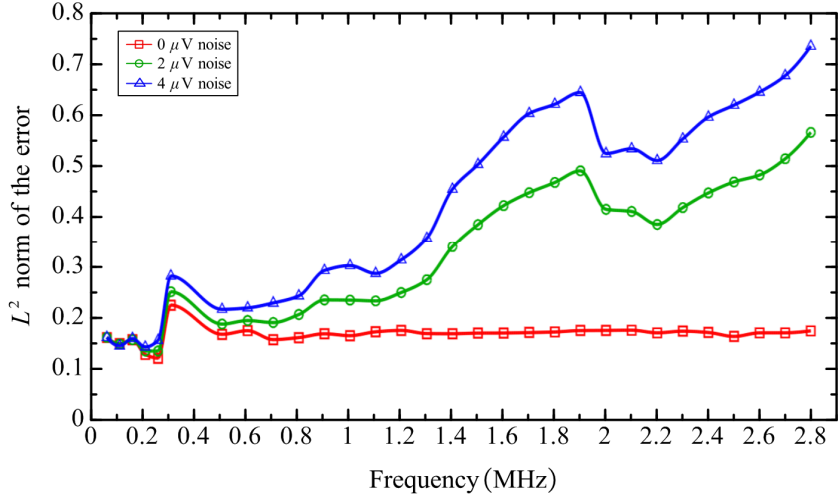

**FIG 9**. $L^2$ norm of the error versus excitation frequency from 0.05 to 2.8 MHz with different noise levels (0, 2 and 4 μV).

The corresponding reconstructed errors are listed in table 1. After noise is added, all reconstructed errors increase. The MAE signals for frequencies below 0.30 MHz are less sensitive to noise. The errors are 0.4421 and 0.6304 with 2 and 4 μV noise for 2.20 MHz, which means the it is difficult to recover the conductivity distribution directly in high frequency with noisy data.

**D. 2D reconstructed conductivity image with B-scan mode**

For Model II, the plane transducer with the centre frequency of 0.25 MHz is used to scan along the $y$ direction. The reconstructed 2D relative conductivity images of Model II with noise-free data are shown in Fig. 11(a). The corresponding conductivity profile (indicated by white dashed line) is presented in Fig. 11(d). The black and red line denote the target and reconstructed conductivity. The reconstructed conductivity images in the presence of 2 μV and 4 μV noise are shown in Fig. 11(b) and (c). The conductivity profiles are displayed in Fig. 11(e) and (f).

The later resolution of plane transducer is very low. Strictly speaking, the conductivity values in Fig. 11 are just averages in each position. For example, the profiles at $y$=-24, -22 and -20 mm are obviously different from that at $y$=-18 mm. Because the main lobe of plane wave does not pass through the anomaly in the middle.

## V. DISCUSSION

In this study, we developed a direct method to retrieve the conductivity distribution of an object. Due to the shortcoming of the measurement formula in the conventional MAET method, we derived the new convolution formula in the form of acoustic velocity and conductivity distribution. We also proposed the non-ideal conductivity model to simulate the practical MAE signal (see supplementary material for complete MAE signal). Moreover, we verified the feasibility of the reconstruction method with numerically simulated data based on two 2D models. Finally, we analyzed several factors influencing the reconstruction error, such as the regularization parameter, ultrasound frequency, and noise.

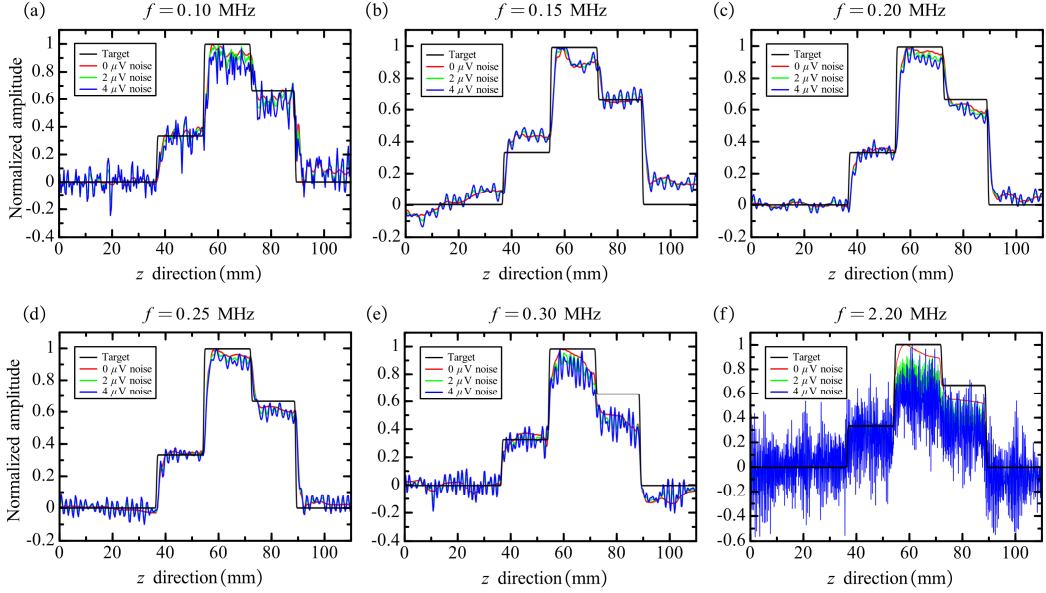

**FIG 10**. Reconstructed conductivity distribution with data of different noise levels for (a) 0.10 MHz, (b) 0.15 MHz, (c) 0.20MHz, (d) 0.25 MHz, (e) 0.30 MHz, and (f) 2.20 MHz.

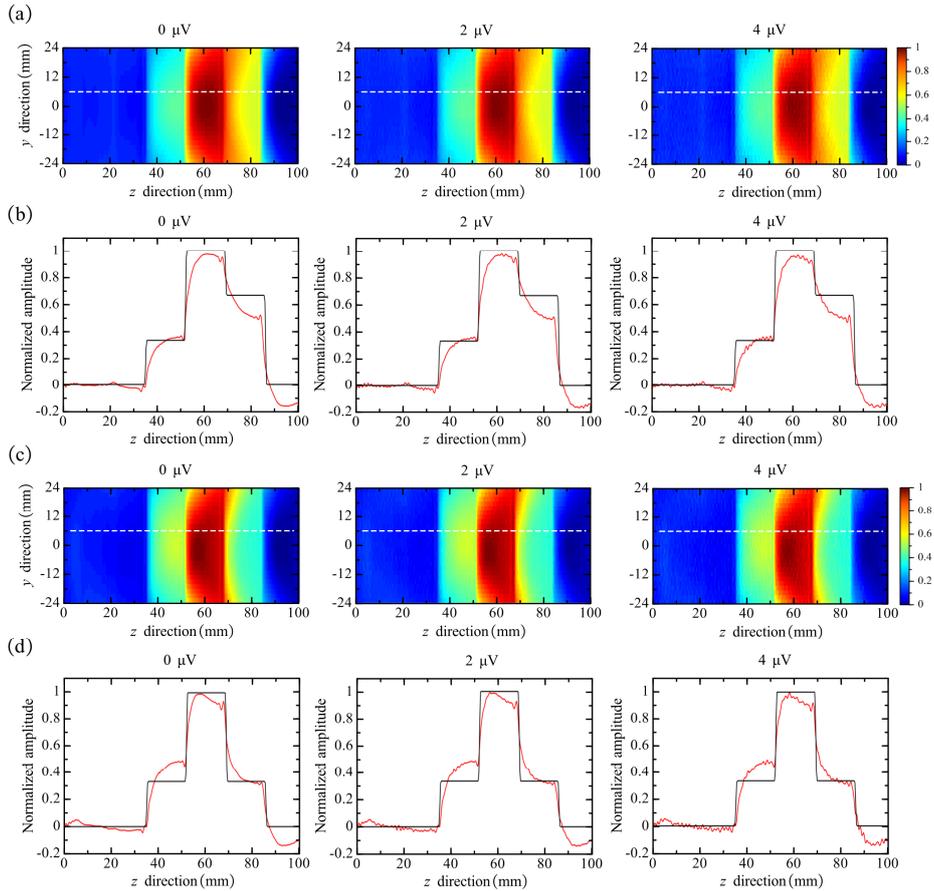

**FIG. 11**. Reconstructed conductivity images for (a) Model I and (c) Model II with 0 $\mu$V noise (left), 2 $\mu$V noise (middle) and 4 $\mu$V noise (right); reconstructed conductivity profile of (b) Model I and (d) Model II with: 0 $\mu$V noise (left), 2 $\mu$V noise (middle) and 4 $\mu$V noise (right); The black and red line denote the target and reconstructed conductivity.

In the conventional MAET method, the MAE signal is directly used to image the conductivity interface. Although the imaging parameter is related to the conductivity, conventional MAET cannot reconstruct the conductivity distribution. In other words, the spatial distribution of conductivity between two interfaces cannot be obtained. Moreover, the waveform of acoustic pressure is not separated from the measured MAE signal, so the conductivity interface image is also called the conductivity parameter image by some studies [36]. To obtain images with high spatial resolution, a higher frequency, such as 2.20 MHz or 2.50 MHz is typically used in the conventional method [27, 34]. In this study, we achieved the goal of reconstructing the conductivity distribution. Moreover, we proved that although the MAE wave packets were aliased with low-frequency excited ultrasound, the conductivity distribution was still recovered by the deconvolution, which meant that the spatial resolution in MAET was not influenced by the duration of excitation ultrasound.

Zhou *et al.* [37] proposed a scheme to recover the conductivity distribution in 1D. Combining the polarity and amplitude information of the deconvoluted signal, Dirac delta functions were introduced manually, and the 1D conductivity distribution was subsequently obtained by utilizing the numerical integral. Using the framework proposed in this study, we can explain the deconvoluted signal in Zhou *et al.* [37]. For the numerical study, only the high-frequency conductivity component was included in their MAE voltage because they performed the convolution of the conductivity gradient and system function. For the conductivity gradient model, the low-frequency component was lost due to the numerical differential.

In this study, we calculated the MAE voltage, including the low-frequency component by using the finite element software COMSOL Multiphysics, which built the model based on the partial differential equation (PDE). A similar MAE voltage with the low-frequency component was also obtained with MATLAB, which built the model based on the numerical calculation of convolution. Consequently, this proved that the measurement formula in the form of convolution was correct, and the low-frequency component related to conductivity existed in the practical experiment.

We found that even when the frequency of ultrasound was low and its duration was long, the conductivity could still be reconstructed. In other words, the spatial resolution was not influenced by the frequency of ultrasound (but the transition band of reconstructed conductivity was influenced by the frequency). Thus, we can use a low-frequency transducer (below 0.5 MHz) to excite the object. We also found that if the range of the transducer did not cover the effective frequency of the object, a rare low-frequency component was included in the MAE voltage. The low-frequency component recovered by the deconvolution was easily contaminated by the measurement noise. One could increase the value of the regularization parameter, but this would lead to the loss of conductivity information. Eventually, only conductivity including the high-frequency component could be only reconstructed (the low-frequency component is lost). In addition, in practice, the power amplifier and the preamplifier also had limited bandwidth, which increased the difficulty of reconstructing the conductivity distribution.

This study had some limitations that should be solved in the future. The deconvolution method was easily contaminated by noise. If the shrinkage increased to compress the noise, the signal was also compressed. In practice, the acoustic velocity along *y* direction was not uniform. The acoustic field should be considered carefully. Another was that the six-cycle sinusoidal burst was used in this study, but the bandwidth of the sinusoidal burst was narrow, which led to the loss of low- and high-frequency components. The ultrasound coded technique with wide bandwidth could be used in the future.

## ACKNOWLEDGEMENTS

This work was supported by the National Natural Science Foundation of China (Grant Nos. 81871429, 91859122, 62071310, and 61971289), Shenzhen Fundamental Research Project (JCYJ20170412111316339), Shenzhen Science and Technology Project (JCYJ20170817171836611), and Shenzhen-Hong Kong Institute of Brain Science-Shenzhen Fundamental Research Institutions.